\documentclass[12pt]{article}
\usepackage[latin1]{inputenc}
\usepackage{amsmath}
\usepackage{amsfonts}
\usepackage{amssymb}
\usepackage{graphicx}
\usepackage{geometry}
\usepackage{hyperref}
\usepackage[english]{babel}
\usepackage[bottom]{footmisc}
\title{Graphic displays of MLB pitching mechanics and its evolutions in PITCHf/x data.}
\author{Fushing  Hsieh\footnotemark, Kevin Fujii\footnotemark[1], Tania Roy\footnotemark[1],  Cho-Jui Hsieh\footnotemark[1] \footnotemark[2], Brenda McCowan\footnotemark[3]
}
\date{}
\begin{document}
\maketitle

\begin{abstract}
	Systemic and idiosyncratic patterns in pitching mechanics of 24 top starting pitchers in Major League Baseball (MLB) are extracted and discovered from PITCHf/x database. These evolving patterns across different pitchers or seasons are represented through three exclusively developed graphic displays. Understanding on such patterned evolutions will be beneficial for pitchers' wellbeing in signaling potential injury, and will be critical for expert knowledge in comparing pitchers. Based on data-driven computing, a universal composition of patterns is identified on all pitchers' mutual conditional entropy matrices. The first graphic display reveals that this universality accommodates physical laws as well as systemic characteristics of pitching mechanics. Such visible characters point to large scale factors for differentiating between distinct clusters of pitchers, and simultaneously lead to detailed factors for comparing individual pitchers. The second graphic display shows choices of features that are able to express a pitcher's season-by-season pitching contents via a series of 3(+2)D point-cloud geometries. The third graphic display exhibits exquisitely a pitcher's idiosyncratic pattern-information of pitching across seasons by demonstrating all his pitch-subtype evolutions. These heatmap-based graphic displays are platforms for visualizing and understanding pitching mechanics.
\end{abstract}
\footnotetext[1]{Department of Statistics, UC Davis} \footnotetext[2]{Department of Computer Science, UC Davis}
\footnotetext[3]{School of Veterinary Medicine, UC Davis}
\section{Introduction}
Both 2017 Cy Young Awards in Major League Baseball (MLB) have coincidentally gone to two previous awardees: Corey Kluber (winner in 2014) of the Cleveland Indians for American League, and Max Scherzer (winner in 2013 and 2016) of the Washington Nationals for National League. Winning multiple Cy Young awards is not rare in MLB history, but is not prevalent at all among MLB pitchers.
This 2017 coincidence is far from implying that MLB pitchers all have steady pitching careers. In fact pitchers' careers mostly go up-and-down from season to season. For instance, among the 24 starting pitchers, who were selected for being one of top-3 candidates for this award through 2012 to 2017 seasons, only Clayton Kershaw is a two-time winner (2013 and 2014). Indeed only three more pitchers: Justin Verlander, David Price and Adam Wainwright, have received top-3 candidacies more than once.

If being one of the top-3 candidates is taken as being at the peak of a pitcher's career, then a natural question is: what characterizes pitching mechanics that only enables a few pitchers to stay at their peaks, and what makes majority of elite pitchers' peaks come and go? These characteristics are at best elusive in this sport. On one hand, aerodynamics of pitching is no doubt primarily governed by Newton's law of gravity and Magnus effect on spin-vs-speed interactions, see Briggs (1959) \cite{briggslyman1959}. It is also critically influenced by biomechanical parameters: like pitcher's physical power and arm strength; and by many non-mechanical parameters: like baseball's skin surface, pitcher's finger conditions and weather conditions. These laws, effects and parameters combine to crucially influence each every pitch's 60 feet 6 inches(18.39m) journey that lasts only 0.5 second or less to go from pitcher's mound to home plate. On top of such a convoluted complexity, a starting pitcher typically has three to five pitch-types in his pitching repertoire, and each pitch-type contains several subtypes. So one explicit and detailed aerodynamic formula for what a pitcher's pitching mechanics looks like is unrealistic, if not unimaginable.

On the other hand, the MLB is at its state of completely missing key information and understanding of pitching mechanics of its pitchers. This dire state is detrimental to this 150-year-old sport. Being unable to explicitly reveal the complexity of pitching mechanics, which is the heart of the game, not only diminishes the basis to excite young people, but also causes many critical issues of pitcher's pitching well-being can't be addressed and dealt with in time. Issues like: Are all pitch-types delivered by a pitcher's in tune? Are there signs of injury?  These natural and serious issues are ought to have proper answers after each game. But they are not.

One primary reason behind lacking rigorous scientific studies on MLB pitching to offer pertinent information and relevant understanding is no availability of proper data. Baseball data has been summarized in the format of a box-score, which is at least as old as the sport. In a box-score out of a game, the amount of information related to pitching mechanics is next to nothing.
However, a revolutionary theme of pitching data collection was finally completed in MLB in 2008.  Such a theme has never been seen before.  Through 2006-2008, MLB and Sportvision, a TV broadcast effects company based in Mountain View, California, installed cameras in all 30 MLB stadiums to track each pitch in every MLB game.

Approximately 20 images are acquired during a baseball's flight from a pitcher's hand to home base, see Mike Fast (2010) \cite{mikefast2010}. These images are used to reconstruct 21 features on various speeds, accelerations, curving and spinning characteristics and coordinates of its release point. Such a database, named PITCHf/x, is now made available by Sportvision to Major League Baseball Advanced Media (MLBAM) to merge together with traditional manual recording of Umpire's calls: strike, ball; and batter's results: hit, home-run, strike-out,etc. Sportvision more recently also created databases, HITf/x and FIELDf/x, to record batted baseballs and all players' movements in the field.

Starting from 2015, the Sportvision's system has been further upgraded by MLBAM's Statcast tracking system based on Trackman's Doppler radar system and ChyronHego's video system. By combining the radar and video technologies, data of all pitched and batted baseballs and all players' movements is recorded in databases. Such a PITCHf/x database is not a data set per se: not only because of its largeness in size, but because of its entire information contents being just too diverse and too vast to be covered and reported in one single paper. In fact the goal of computing upon this database is set to build various platforms, on which all people can get access to their own information, and to foster their own understanding, and then to collectively produce fundamental and brand new insights into the sport. By now this database has been publicly available for almost 10 years. It is the right time that we build such platforms. Hopefully people can explore and look into all pitcher's pitching mechanics, and see this sport from new perspectives. With the aforementioned goal in mind, we focus this paper on developing data-driven computational techniques to extract pertinent information of pitching mechanics based on PITCHf/x database. Then we build graphic displays as platforms for representing computed information in both systemic and idiosyncratic fashions.

From systemic perspective, a mutual conditional entropy matrix, which reveals the dependency among all 21 features based on combinatorial information theory, is computed across all seasons of all 24 starting pitchers. We discover a universal composition of patterned blocks upon all matrices, which is a structural manifestation of aerodynamics of baseball pitching. This universality also frames a graphic display that contains interacting relations capable of characterizing large and small scales distinctions among clusters of pitchers.

For the idiosyncratic perspective, this universality also implies which features can exhibit evolving characteristics as a quick graphic display of a pitcher's pitching mechanics through visible 3(+2)D geometries (size and color). For detailed evolutions within all pitch-types in a pitcher's repertoire, a graphic display of categorical likelihood processes of all identified pitch-subtypes along their pitch-by-pitch temporal coordinates is constructed. Such a graphic display offers complete information of evolutions of a pitcher's pitching mechanics across all seasons. Hence it is a platform for seeking what characters are contributing to his career peaks, and what characters are causing up-and-down along a pitcher's career path. It can also reveal emergent patterns for potential injures or being out-of-tune. In similar fashion, we are able to compare several pitchers' pitching mechanics in detail by having one of them as a baseline.

In this big-data era, our data-driven computing and graphic displays can transform PITCHf/x database into ``modern scientific platforms" that are sophisticated enough at this computer age in order to attract the young minds, and at the same time are insightful enough to educate the young brains as well as to challenge the old wisdoms.

\section{Materials and Methods}
\subsection{Aerodynamics of baseball pitching and its characteristic features}
The aerodynamics of a baseball pitch in reality is rather complex, since it involves several physical laws and many natural and bio-mechanical parameters. It is not only too complex to describe in full, but also too expensive to record and store in fine detail. Further a starting pitcher typically throws about 100 pitches per game. With four or five starting pitchers rotating throughout the 162 games in a season, a pitcher can produce above 3000 pitches a year, not including play-off games. So, in order to store coherent information of a pitcher's pitch-by-pitch aerodynamics, a private company, Sportvision contracted by MLB, has chosen 21 features out of each pitch's flight from the moment the baseball leaving the pitcher's hand to the moment passing through home plate. Ideally these 21 features across approximately 3000 pitches should contain each pitcher's pitching mechanics.

It is well known that several key characteristics of pitching mechanics are governed by a physical law, called the Magnus effect. This effect primarily prescribes the complicate nonlinear interacting relations between the baseball's traveling speed and spin under influences derived from forces of gravity and air resistance. The starting speed (``Start\textunderscore speed") is measured when the ball is at the point 50 fts away from the home plate, which is very close to the release point ``$(x_0, y_0, z_0)$" of a pitch.  The spin direction (``spin\textunderscore   dir") and spin rate (``spin\textunderscore   rate") are attributed to complex factors, such as, how a baseball is held in a pitcher's hand; how it is released; how pitcher's fingers touch and slide against the surface and seams of a baseball. Particularly the ways of holding and sliding against baseball seams are responsible for various pitch-types contained in a pitcher's repertoire. Indeed a MLB pitcher can be seen as an ``expert" manipulator of speed and spin of a baseball.

This effect of a baseball pitch is briefly described as follows. With a high enough ``start\textunderscore   speed", spin tends to stabilize a baseball's trajectory linearly even against the air resistance. That is, spin makes the effects of baseball's uneven surface insignificant at a high speed.  That is why a 100 mph fastball, typically having high spin rate, looks straight. In sharp contrast, a knuckleball, typically having zero spin, has a rather unpredictable trajectory when it arrives at the home plate. It often causes an experienced catcher to miss the catch. When a baseball's traveling speed is gradually reduced, the Magnus effect begins to show on its trajectory. The backspin fastball will go against the gravity and move upward when arriving at the home plate. In contrast, the topspin curveball will go downward and drop more than the effect caused by gravity.  So topspin and backspin cause positive and negative vertical movements, which are measured by a feature at the home plate and denoted by ``pfx\textunderscore   z". Thus this feature has a high association with ``start\textunderscore   speed" for pitchers, who has the high speed fastball as his chief pitch-type in his repertoire, than for pitchers, who doesn't.

The feature ``pfx\textunderscore   z" is also associated with features related to how a baseball trajectory curves. A baseball trajectory from release point to the home plate is coupled with two straight lines: the tangent line at the release point ``$(x_0, y_0, z_0)$" and the line links the release point and the trajectory's end point. The angle between these two lines is termed ``break-angle", while the maximum distance between the baseball trajectory and the second straight line is called and denoted as ``break\textunderscore length".  Therefore the three features: ``pfx\textunderscore z", ``break-angle" and ``break\textunderscore length", are highly associated with each other.

It is also intuitive that this ``break\textunderscore angle" could be critically affected when the baseball trajectory indeed swerves to the right or left sides of home plate. Such horizontal swerving is caused by side-spin, which is often induced in pitch-type: like slider (SL) or change-up(CH). So side-spin makes the baseball to swerve to the corresponding side and produces the horizontal movement, which is denoted as ``pfx\textunderscore   x".  Therefore the three features: ``pfx\textunderscore   x", ``break-angle" and ``spin\textunderscore   dir", are mechanistically associated. A skilled pitcher usually makes use of certain ``spin\textunderscore   dir" to produce a large ``pfx\textunderscore   x" value to deal with a batter. That is why a right-handed pitcher can be more effectively to against a right-handed batter, but less effective to left-handed one. A right-handed pitcher likely throws pitches with large horizontal movements to the left. Such an effect explains why proportions of left-handed batters and pitchers in MLB are as high as $40\%$, which is twice the proportion of left-handed people in general population.

In summary, the aerodynamics or pitching mechanics of a baseball pitch are basically governed by six features: ``start\textunderscore speed", ``spin-rate", ``spin-dir", ``pfx\textunderscore z", ``break\textunderscore angle" and ``break\textunderscore length" in an intricate and connected fashion. The remaining 15 features are either highly associated with these six features individually and collectively, such as  ``end\textunderscore speed"  and three directions of speeds and accelerations at the lease point, named ``vx0, vy0,vz0" and ``ax, ay, az", respectively, or play only auxiliary roles, like ``break\textunderscore y"  and ``($x_0, z_0$)".

\subsection{Graphic displays as Representations of knowledge in PITCHf/x data}
The object of data-driven computing here is information of pitching mechanics. The objective of such computing is the understanding of the information contained in PITCHf/x data. To stimulate such understanding we employ graphic displays as representations of computed systemic and idiosyncratic pattern information of MLB pitching mechanics. For systemic information, our graphic display is composed of a clustering composition of pitchers featured with computed patterns on: how some pitchers aggregate, some set apart, and why one pitcher appears on different branches over different years. For idiosyncratic information, our graphic display represents the evolving processes of all pitching-subtypes on: when it goes extinct; what are remaining stable; how only a few are created or re-borne. Also for idiosyncratic information, the graphic display via serial 3(+2)D geometries are designed to reveal visible characters of pitching mechanics through key features across a series of seasons.

The guiding principle of employing graphic displays is to appeal to human's formidable visual and mental processing capabilities. In this computer era, our brain coupled with visual sensory systems might be still the most efficient device for recognizing and organizing patterns into understanding and knowledge, see also Grenander and Miller (1994) \cite{GrenanderMiller94}.

\section{Computing Methods}
\subsection{Possibly-gapped Histogram for categorical renormalization}
Among 21 features, there are drastic different measurement units involved. We explore potential non-linear associations among these features based on combinatorial information theory. For this application, we transform each real valued feature into a categorical one as a way of renormalization. This categorization also prepare us for calculations of mutual conditional entropy. The real-to-categorical transformation is performed by applying Analysis of Histogram (ANOHT) algorithm, which builds a possibly-gapped histogram upon 1D real valued data set. So one bin is one category.

Such a possibly-gapped histogram is originally designed to reveal pattern information contained in 1D data via uniformity within all bins of various sizes. And a gap is an extreme form of uniformity. The validity of such a histogram is visible through its corresponding possibly-gapped piecewise-linear approximation to its empirical distribution, see details in Fushing and Roy (2017) \cite{hsiehroy}.

\subsection{Mutual conditional Entropy based on combinatorial information theory}
A possibly-gapped histogram transforms the 1D feature into an (ordinal) categorical variable upon the ensemble of pitch-IDs. Two features, for instance, say, break\textunderscore   length and start\textunderscore   speed, will give rise to a bivariate categorical variable upon the same pitch-ID ensemble. This bivariate categorical variable can be also expressed via a contingency table, for instance, with break\textunderscore   length's categories being arranged along the row axis and start\textunderscore   speed's categories on the column axis.

Within a given category of break\textunderscore   length, equivalently being fixed on one row of the contingency table, we evaluate Shannon entropy with respect to categories of the start\textunderscore   speed. Further we calculate the ratio of this Shannon entropy with respect to the overall Shannon entropy of start\textunderscore   speed. This ratio is the relative local conditional entropy of start\textunderscore   speed given Break\textunderscore   length pertaining to a specific category. It is taken as a directed local association from start\textunderscore   speed  to break\textunderscore   length, that is, this directed local association is meant to evaluate the degree of exclusiveness of categories of start\textunderscore   speed within a specific category of break\textunderscore   length.

The global start\textunderscore   speed to break\textunderscore   length association will be calculated by a weighted sum of all local ones with weighting being the proportions of break\textunderscore   length's category sizes. Finally the bi-directed global association of break\textunderscore   length to start\textunderscore   speed based on their relative mutual condition entropy is the average of global conditional entropy from break\textunderscore   length to start\textunderscore   speed and global conditional entropy from start\textunderscore   speed to break\textunderscore   length. This is how the non-linearity is accommodated in a measure of association between start\textunderscore   speed and break\textunderscore   length. The most important merit of this association is that, as an effective summarizing statistics of the contingency table, it conveys the authentic dependency between start\textunderscore   speed and break\textunderscore   length. A pictorial illustration of this mutual conditional entropy can be found in an introductory paper Fushing et al. (2017) \cite{hsiehshanyu}.

Likewise a 21X21 matrix of mutual conditional entropy is constructed, see examples in Figure \ref{fig:figure1}. This matrix will be taken as a platform for collectively revealing the dependency structures among these 21 features.

\subsection{Synergistic feature-groups and backbone of dependency}
The 21X21 mutual conditional entropy matrix is taken as a distance matrix among the 21 features. Upon this distance matrix, Hierarchical Clustering (HC) algorithm is applied to construct a HC tree on this ensemble of features. By superimposing the HC tree on the matrix's row and column axes, multi-scale block-patterns are framed and revealed, see examples in Figure \ref{fig:figure1}.

Each block along the diagonal corresponds to a cluster of features by having uniformly low mutual conditional entropies (high associations). Such a feature-cluster is called a synergistic feature-group. Thus a series of diagonal blocks will signal a series of synergistic feature-groups that potentially serve as one chain of mechanistic dependency structure among these 21 features.

\subsection{Data Mechanics for systemic characteristics of pitching mechanics}
To explore systemic characters of all involving pitcher-seasons, we convert the upper triangular part of 21X21 mutual entropy matrix into a 210-dimensional vector. By stacking all pitcher-year's 210-dim vectors together, a rectangle matrix with 210 columns is constructed.

Next, for computational simplicity and costs, we adapt hierarchical clustering algorithm on building a tree on row axis of pitcher/year and another tree on column axes. These two trees are supposed to be coupled because of interacting relations between pitcher-seasons on row-axis and pairwise mutual conditional entropies on column-axis. We then construct these two trees in an iterative fashion: 1) starting from building a HC tree on column axis with Euclidean distance; 2) define a new distance between row vectors as the sum of 210 dimensional Euclidean distance and another Euclidean distance of extra dimensions based on a clustering composition of cluster-averages with respect to a selected tree-level of the HC on column axis, and the build a HC tree on row axis; 3) the distance between column vectors is updated in the same manner with respect to the tree derived in Step 2. A distance being updated with respect to a tree structure is designed as a way of enhancing the universal block-patterns found in the mutual conditional entropy matrix. After iterating once or twice, the resultant HC tree on the row axis gives rise to multiple scales of similarity among pitcher-seasons, while the block patterns on the matrix's rectangle lattice, (or heatmap), reveal the systemic characteristics for each cluster of pitcher-season, see examples in in Figure \ref{fig:figure3}.  Likewise computations can be performed on a sub-matrix of the 21X21 matrix.

This iterative HC algorithm mimics the newly developed computing paradigm called Data Mechanics, see Fushing and Chen (2014) \cite{hsiehchen} on binary rectangular matrix and Fushing, et al. (2015) \cite{Fushingchihhsin2015interface} on real-value matrix. In Data Mechanics, Ultrametric trees are derived based on a data-driven algorithm called Data Cloud Geometry (DCG), see Fushing and McAssey (2010) \cite{fushing2010} and Fushing, et al. (2013) \cite{fushingwangvanderwaalmccowankoehl13}. DCG was designed to extract the authentic tree information contained in the data. It is contrasting with the man-made artificial binary splitting on each HC tree branch. A DCG tree will require a higher computing cost than HC tree will do. However the key step of Data Mechanics is the iterative procedure. Hence the iterative procedure based on HC algorithm is still termed Data Mechanics (DM).

{\bf [Data Mechanics for idiosyncratic characteristics: Pitch-subtype evolution.]}
Again Data Mechanics is applied to explore a pitcher's idiosyncratic characteristics. This exploration into pitching mechanics is pitch-type specific. Consider a pitch-type specific rectangular matrix with 21 or a set of selected features being arranged on column axis and all pitches across a series of consecutive seasons on row axis. By applying DM, a 6 cluster-level of the clustering tree on the row axis is selected to define 6 pitch-subtypes, so each pitch has a pitch-subtype-ID. At the same time, the clustering tree on column-axis reveal the visible characters that define each pitch-subtype in an explicit compositional fashion, see examples in Figure \ref{fig:figure6}.

Among all involving seasons, a subset is selected to serve as the baseline seasons, in which the pitcher of interest is considered being healthy.  Then the 6 subtype-specific proportions of pitches belonging to the baseline seasons are calculated. Each proportion is termed a likelihood of a pitch pertaining to its specific pitch-subtype. Together they form a generically called categorical pattern distribution.

Accordingly such a categorical pattern distribution depicts how a pitch relates to pitching mechanics of the baseline seasons. A large likelihood value indicates that this pitcher continues to pitch a high potential subtype of the baseline seasons. While a zero, or an extremely small likelihood value indicates a brand new subtype being created outside of the pitching repertoire employed in the baseline seasons. Therefore, when a pitch is displayed with its pitch-subtype-ID and likelihood value along with its coordinate on the temporal axis, a graphic display of evolution of pitch-subtype is constructed for a specific pitch-type.

\section{Results}
\subsection{Universality}
\begin{figure}[hbtp]
	\centering
	\includegraphics[scale=0.8]{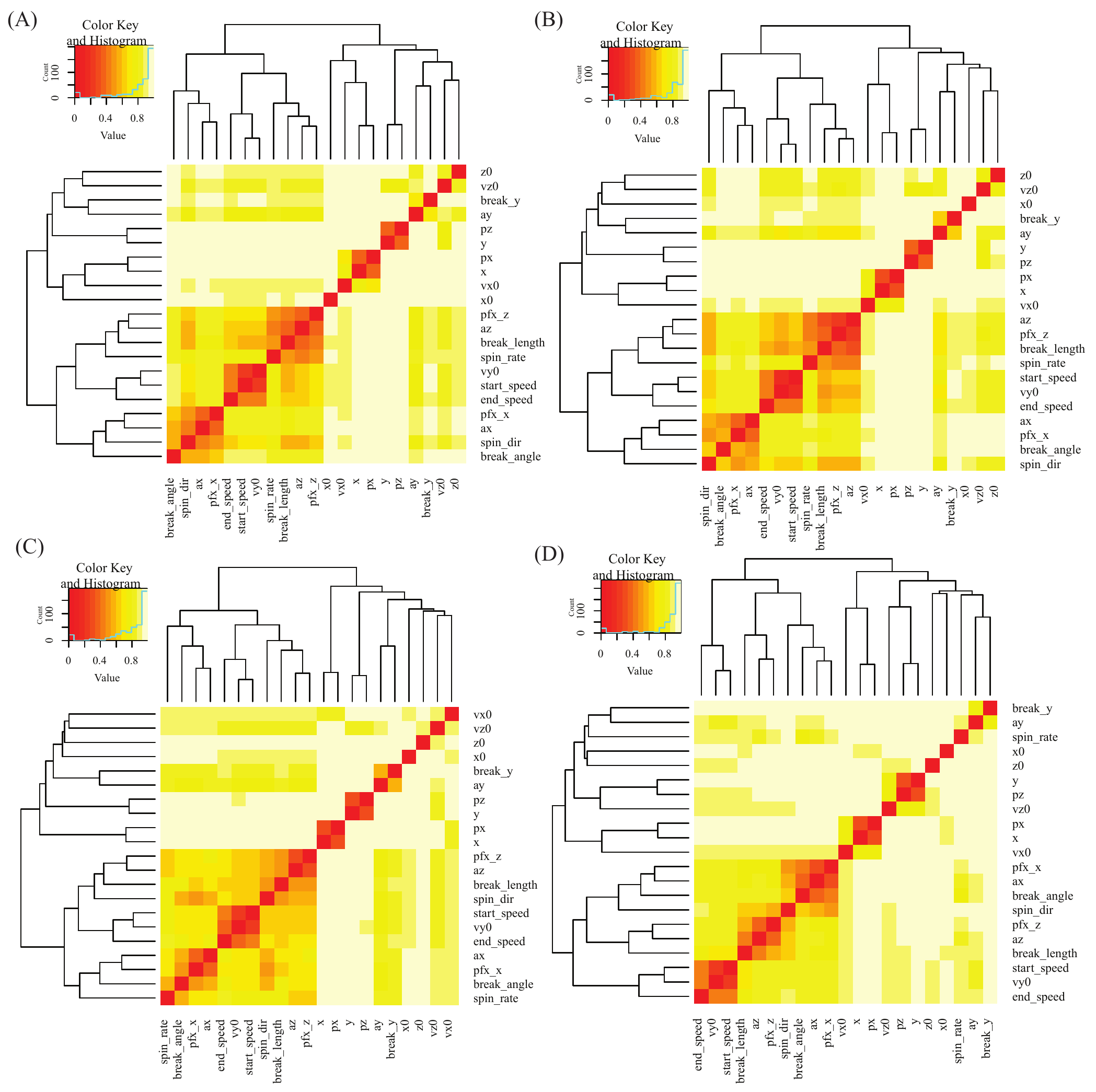}
	\caption{Mutual conditional entropy matrices: (A) Clayton Kershaw at 2013 season; (B) Clayton Kershaw 2014; (C) Justin Verlander 2016; (D) Kyle Hendricks 2016.}
	\label{fig:figure1}
\end{figure}

For all 24 pitchers, their 108 pitcher-seasonal mutual conditional entropy matrices universally show a composition of two serial blocks on two distinct scales along the diagonal: the large scale one is a single 11x11 sub-matrix; the small scale ones are: one 4x4, two 3x3 and one 1x1 submatrices contained within the 11x11block, and two 3X3 and one 2x2 outside the large block, see Figure \ref{fig:figure1}. (All 24 pitchers' evolving mutual conditional entropy matrices across a series of seasons are reported in the DM website \url{https://www.dm-mlbpitching.com}.) 


Each of small scale block reveals how one small set of synergistic features works as one mechanical component within the pitching mechanics. For instance, the four mechanical components: \{start\textunderscore speed, end\textunderscore speed, vy0\}, \{pfx\textunderscore z, az, break\textunderscore length\}, \{break\textunderscore angle, pfx\textunderscore x, ax, spin\textunderscore dir\} and \{spin\textunderscore rate\}, are evident parts of pitching aerodynamics. The mechanism via synergistic feature group:\{pfx\textunderscore z, az, break\textunderscore length\} depicts how and how much the baseball trajectory would curve in the vertical direction. The synergistic feature group:\{break\textunderscore angle, pfx\textunderscore x, ax, spin\textunderscore dir\} depicts how and how much a baseball trajectory would move in the horizontal directions: left or right. These two mechanical components are closely associating with Magnus effect. In contrast, the 11x11 block depicts how the four mechanical components collectively work together to form the aerodynamics commonly shared by all pitching mechanics of the 24 pitchers, while the remaining 10 features outside the 11x11 block are relatively low in overall association because they are more related to pitchers' idiosyncratic pitching gestures than pitching mechanics per se.

\begin{figure}[hbtp]
	\centering
	\includegraphics[width=0.7\linewidth]{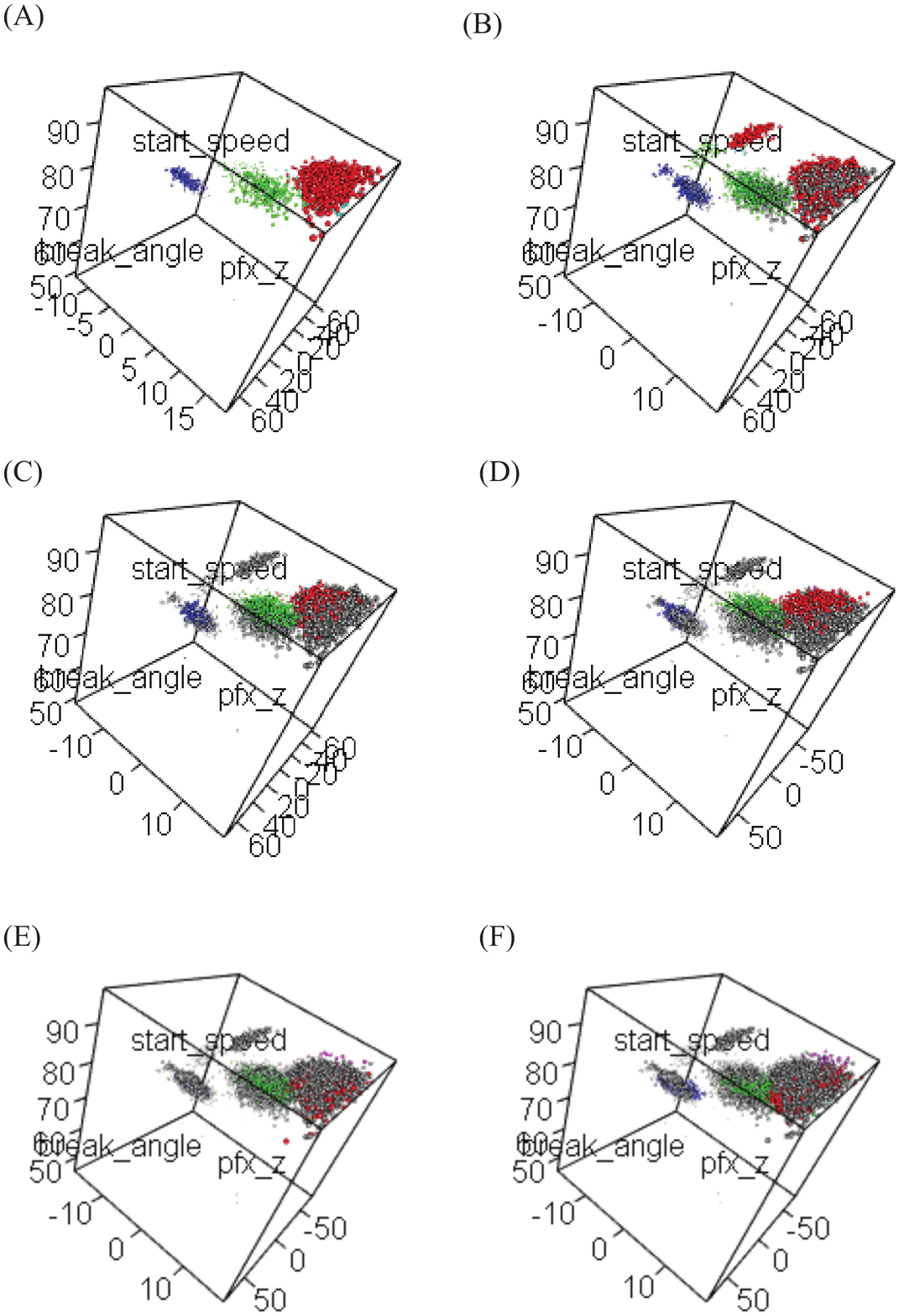}
	\caption{Clayton Kershaw's progressive data cloud geometries of pitch-types from season 2012-2017. (A) 2012; (B) 2013 with 2012 as baseline; (C) 2014 with 2012-13 as baseline; (D) 2015 with 2012-14 as baseline; (E) 2016 with 2012-15 as baseline; (F) 2017 with 2012-16 as baseline. Color-coding: FF in Red; FT in Orange: SL in Green; CU in Blue.}
	\label{fig:figure2}
\end{figure}

This universality confirms the point-cloud geometry of 21 dimensional features of all pitches from each pitcher within one single season indeed contains all physical laws underlying pitching mechanics. Two important merits are implied by such a universality.
First, essential features can be selected out of the fine scale blocks to facilitate 3(+2)D geometric display of a pitcher's seasonal pitching mechanics and its evolution across a series of seasons, as seen in Figure \ref{fig:figure2}. The three chosen coordinates are: \{start\textunderscore speed, pfx\textunderscore z, break\textunderscore angle\} plus two extra-dimensions of color-coded pitch-type on each focus season and \{spin-rate\} for sizes of balls. This choice of coordinates provides a visualization of separating pitch-types into cloud-like clusters.

Further, as visualizing color-coded pitches (in ball shape) of current season against previous seasons along the serial geometries, if balls of one color are only seen sparsely within a cloud, then this pitcher's corresponding pitch-type in this focal season is coherent with his in the previous seasons. Such phenomena are consistently seen in Kershaw's pitching mechanics from 2012 through 2017.  On the other hand, if balls of one color are evidently seen aggregating outside the gray-colored cloud, then we should suspect that this pitcher likely being out of tune often in the focal season, if not injured. (See DM website for all 24 pitchers' pitching mechanics evolution through a series of seasonal 3(+2)D animation.)

\begin{figure}
	\centering
	\includegraphics[width=0.96\linewidth]{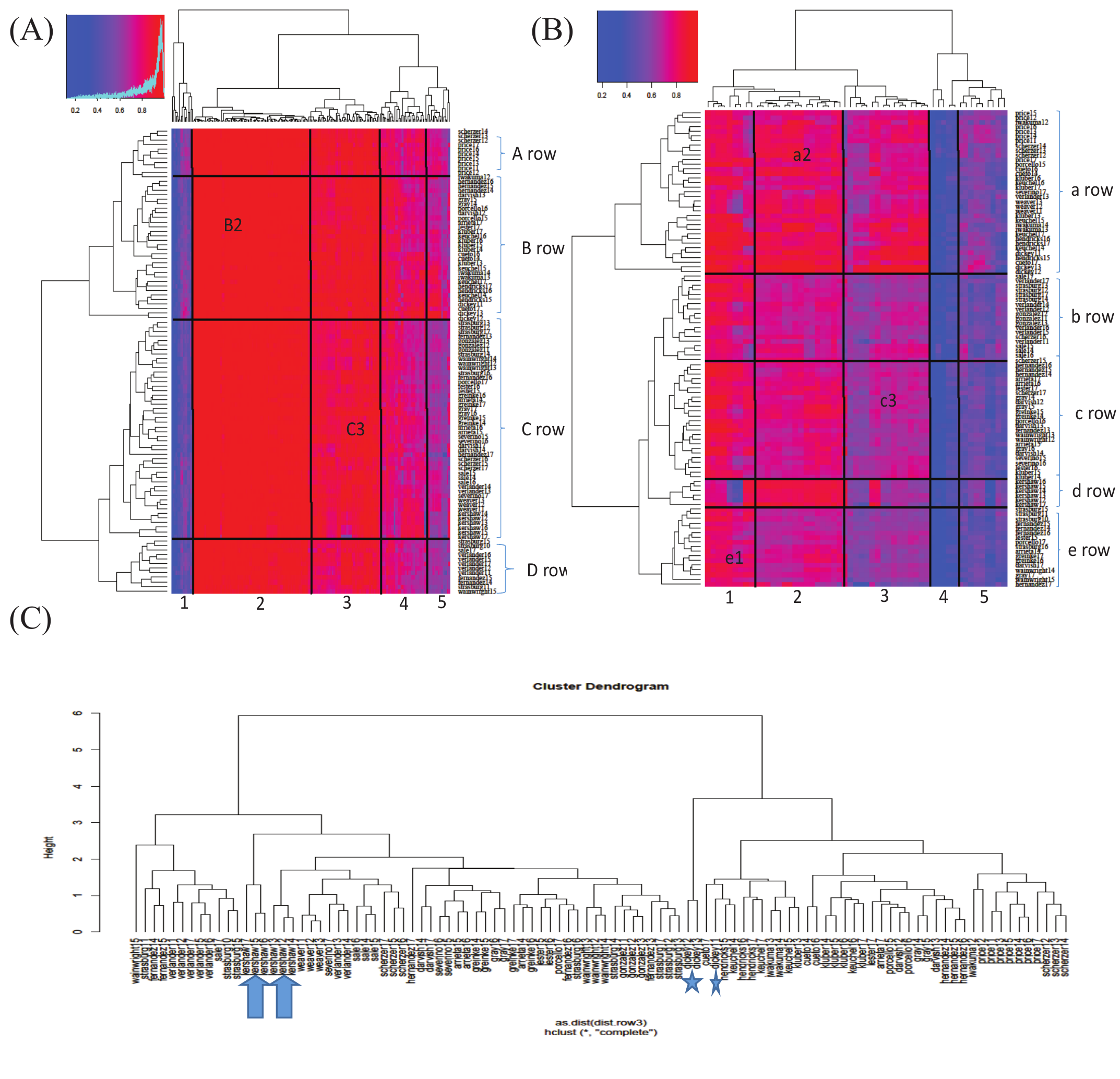}
	\caption{Two versions of heatmap-based systemic patterns: (A) All feature-pairs among 21 features; (B) feature-pairs among 11 selected features; (C) The tree locations of Clayton Kershaw and R. A. Dickey.}
	\label{fig:figure3}
\end{figure}

Secondly, the off-diagonal entries beyond the four small blocks within the 11X11 submatrix collectively characterize how synergistic-group-based mechanical components interact. Such quantified interactions importantly provide information about how individual pitchers' pitching mechanics would differentiate from others. In fact more detailed individual differences in pitching mechanics and biomechanics are summarized in the off-diagonal entries of the whole 21X21 mutual conditional entropy matrix. Two versions of systemic comparisons based on ``pitcher-season" are reported in first two panels of Figure \ref{fig:figure3}. The two heatmaps have 55 (=11x10/2) and 210 columns, respectively.

Each heatmap provides a visible platform for systemic comparisons among ``pitcher-seasons" and for understanding how and why they are similar or distinct. Specifically the marked-blocks, which are framed by the clustering tree of pitchers on the row axis, and another clustering tree of feature-pairs on the column axis, reveal information not only about who is close to and far away from whom, but also, more importantly, about whether a pitcher is close to himself across a series of season. Understanding of information of similarity and distinction can be easy seen through the coloring of the block.

For instance, upon the heatmap in panel (A) of Figure \ref{fig:figure3} pertaining to the 11 features, Kershaw-12 through Kershaw-17 exclusively occupy the ``d-row" branch. This fact means that Kershaw's pitching mechanics is rather steady throughout the 2012-17 seasons from the mechanical aspect pertaining to the 11 features. Also his pitching mechanics is not extremely, but rather different from many pitchers'.  In contrast, Scherzer-12 through Scherzer-17 are found in ``a-row" for 2012-14 seasons, 2016 in ``b-row" and 2015 and 2017 in ``c-row", likewise for Kluber's  2012 through 2017 ,and Verlander's 2011 through 2017 seasons.

However, by adding biomechanical aspects, that is, pertaining to the 21 features, the heatmap in panel (B) and its clustering tree in panel (C) of Figure \ref{fig:figure3} can provide different version of comparison among pitcher-seasons. The 6 seasons of Kershaw are divided into two three-season branches: 2012-14 vs 2015-17 (marked by two upward-pointing arrows). The detailed information attributed to this separation would be seen in the later in this section.

It is also informative to note that a branch consisting of Dickey-12 and -13 stands out as an outlier branch away from the rest of pitcher-season, including his own Dickey-11. The distinct separation can be attributed to the fact that R. A. Dickey had been developing his knuckleball through 2011 (marked by a small star). His knuckleball became very effective in 2012 season (together with 2013 marked by a star), in which he received the Cy Young award. Since the 2012 season, he has been recognized as one of the best knuckleball pitchers in American baseball history. There must be many patterns and related understanding waiting to be discovered from Figure \ref{fig:figure3} alone by devoted experts.

\subsection{Comparing pitchers' with specific feature-pairs}
The heatmaps in Figure \ref{fig:figure3} also guide us to see: how to choose which set of features for comparing a chosen set of pitchers. For instance, three pitchers: Hendricks, Kershaw and Verlander, are to be compared upon four features, \{break\textunderscore length, Spin\textunderscore rate, pfx\textunderscore z, az\}, which constitute three feature-pairs located in the cluster No.1 of feature-pairs in panel (A).

\begin{figure}[hbtp]
	\centering
	\includegraphics[width=1\linewidth]{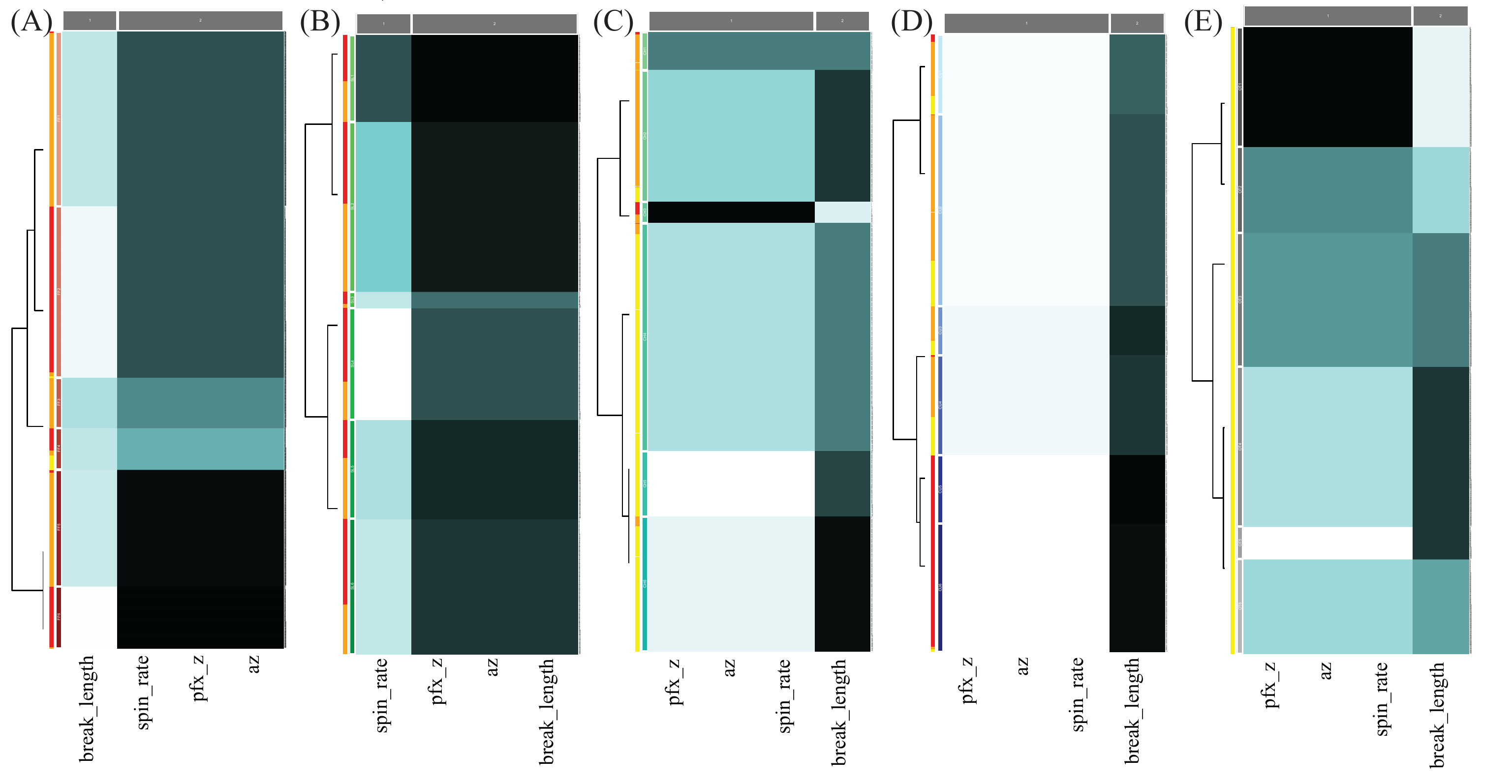}
	\caption{Heatmaps of pitch-type specific comparisons among Clayton Kershaw (baseline), Justin Verlander and Kyle Hendricks Hendricks on 2014-2016 seasons with respect to four features: Break-length, pfx\textunderscore z, az and spin-rate. (A) FF; (B) SL; (C) CH; (D) CU; (E) OT: all pitch-types out of Kershaw's repertoire.}
	\label{fig:figure4}
\end{figure}

Specifically three significant different block values are visible: ``very red" for Hendrick-16 at ``a-row", ``very blue" for Kershaw-16 at ``d-row" and ``intermedium" for Verlander-16 at ``b-row".  The detailed comparison is performed by applying the Data Mechanics on pitch-type specific data matrices having 4 columns for the four features and all pitches of the three pitchers in 2016 season on row-axis. The 5 resultant heatmaps from Data Mechanics computations for four Kershaw's pitch-types and one pooled pitch-type beyond his are shown in Figure \ref{fig:figure4}.

The panel (A) of Figure \ref{fig:figure4} shows the heatmap of fastball (FF), including 2-seam fastballs (FT), of these three pitchers, which are color-coded. Hendricks' fastball pitches (Yellow colored) only show up in subtype cluster FF4 together with Kershaw's (Red colored) and a rather small number of Verlander's (Orange colored). Thus, it is evident that Hendricks' fastball pitches are drastically different from Kershaw's and Verlander's with respect to the four features.

From the heatmap, we also see that this FF4 cluster has rather smaller values in features \{pfx\textunderscore z, az, spin-rate\}. Further, since the three features: \{pfx\textunderscore z, az, break\textunderscore length\}, are highly associative as forming a 3x3 diagonal block within the universality, we understand that Hendricks' fastball pitching mechanics has much smaller vertical acceleration (az) and vertical movement (pfx\textunderscore z) and spin-rate. From outlook of a pitch, his fastball is not as sharp as Kershaw's and Verlander's.

On the same heatmap, we see that Verlander's and Kershaw's fastball pitching mechanics are distinct only on the \{break\textunderscore length\} feature. The three clusters: FF1, FF3 and FF5, are nearly completely dominated by Verlander's, while clusters: FF2 and FF6 nearly completely dominated by Kershaw's. This slight difference in value of break\textunderscore length reveals that Verlander's fastball curves more than Kershaw's fastball.

As for slider (SL) in panel (B) of Figure \ref{fig:figure4}, we see that all clusters are orange and red colored, that is, Kershaw and Verlander have large overlaps on all 6 subtypes of slider. This fact means that their pitching mechanics for slider are rather comparable. Slider is not in Hendricks' pitching repertoire.

In panel (C) for changeup (CH), two clusters: CH1 and CH2 are nearly completely dominated by Verlander, while 3 clusters: CH4, CH5 and CH6, are nearly completely dominated by Hendricks. The smallest cluster: CH3, is shared by Verlander and Kershaw. This rather distinct subtype is in opposite characters of the rest of 5 subtypes.

The curveball(CU) in panel (D) of Figure \ref{fig:figure4} shows a very interesting comparison among these three pitchers. The four subtype-clusters: CU1-CU4, are color-coded with orange and yellow, while two subtype-clusters: CU5 and CU6, are entirely color-coded with red. These 6 curveball subtypes are relatively similar with respect to the features and their corresponding mechanism. The only difference is that Kershaw's curveballs curve more than Verlander's and Hendricks'. The panel (E) of Figure \ref{fig:figure4} shows that Hendricks has certain pitch-types that are not in Kershaw's or Verlander's repertoires.

\begin{figure}[hbtp]
	\centering
	\includegraphics[width=1\linewidth]{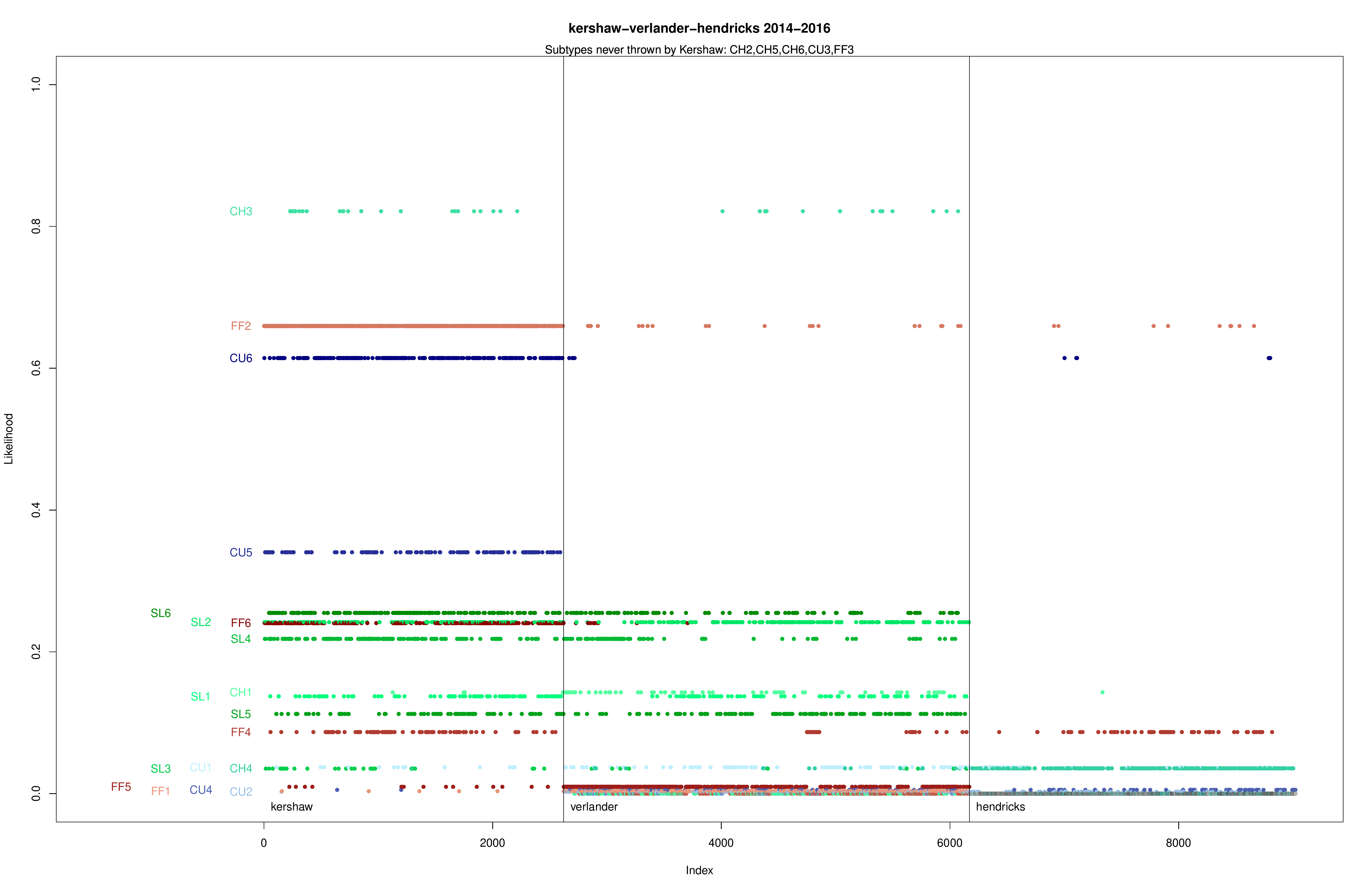}
	\caption{Pitch-subtypes based evolution for comparing three pitchers: Clayton Kershaw (baseline), Justin Verlander and Kyle Hendricks on 2014-2016 seasons with respect to four features: Break-length, pfx\textunderscore z, az and spin-rate.}
	\label{fig:figure5}
\end{figure}

The aforementioned similarity and distinctions among these three pitchers' pitching mechanics from the aspect of the four features: \{break\textunderscore length, Spin\textunderscore rate, pfx\textunderscore z, az\}, are summarized in Figure \ref{fig:figure5}. By taking Kershaw as the baseline for deriving his 5 categorical pattern distributions on the five pitch-types, each pitch then has acquired a pitch-subtype ID, such as FF1, and a likelihood value according to a categorical pattern distribution. Also every pitch from a pitcher has a temporal coordinate. Therefore, when all pitches of a pitcher can be laid out with respect to its ID, likelihood value and temporal coordinate, as such a graphic display is constructed, as shown in Figure \ref{fig:figure5}. This graphic display provides the 4-feature-view of pitching mechanics comparison from the perspective of Kershaw.  Similar viewing perspectives from Verlander and Hendricks can be likewise constructed.

\subsection{Individual pitcher's pitching evolution}
For one individual MLB pitcher, it is critical to be able to see how his pitching mechanics evolves across a series of consecutive seasons, particularly including his career peaks. Here we construct and show Kershaw's pitching mechanics from 2012 through 2017 seasons. There are 5 pitch-types: FF, FT, SL, CH, CU, in his pitching repertoire.  His fastball is subdivided into: 4-seam fastball (FF) and 2-seam fastball (FT).

Data Mechanics computations are applied onto the five ensembles of pitches collected from the 6 seasons. Each pitch's season-ID is color-coded: {red, orange, yellow, green, blue, purple}, from 2012 to 2017 in increasing order. Each pitch is characterized by 13 features: 11 features contained in the universality and \{x0, z0\} the horizontal and vertical coordinates of its releasing point. The reason for including these two features is that they are closely related to pitching gesture at large. Kershaw is known for his rather distinct gesture particularly for his curveball pitching.

\begin{figure}[hbtp]
	\centering
	\includegraphics[width=0.9\linewidth]{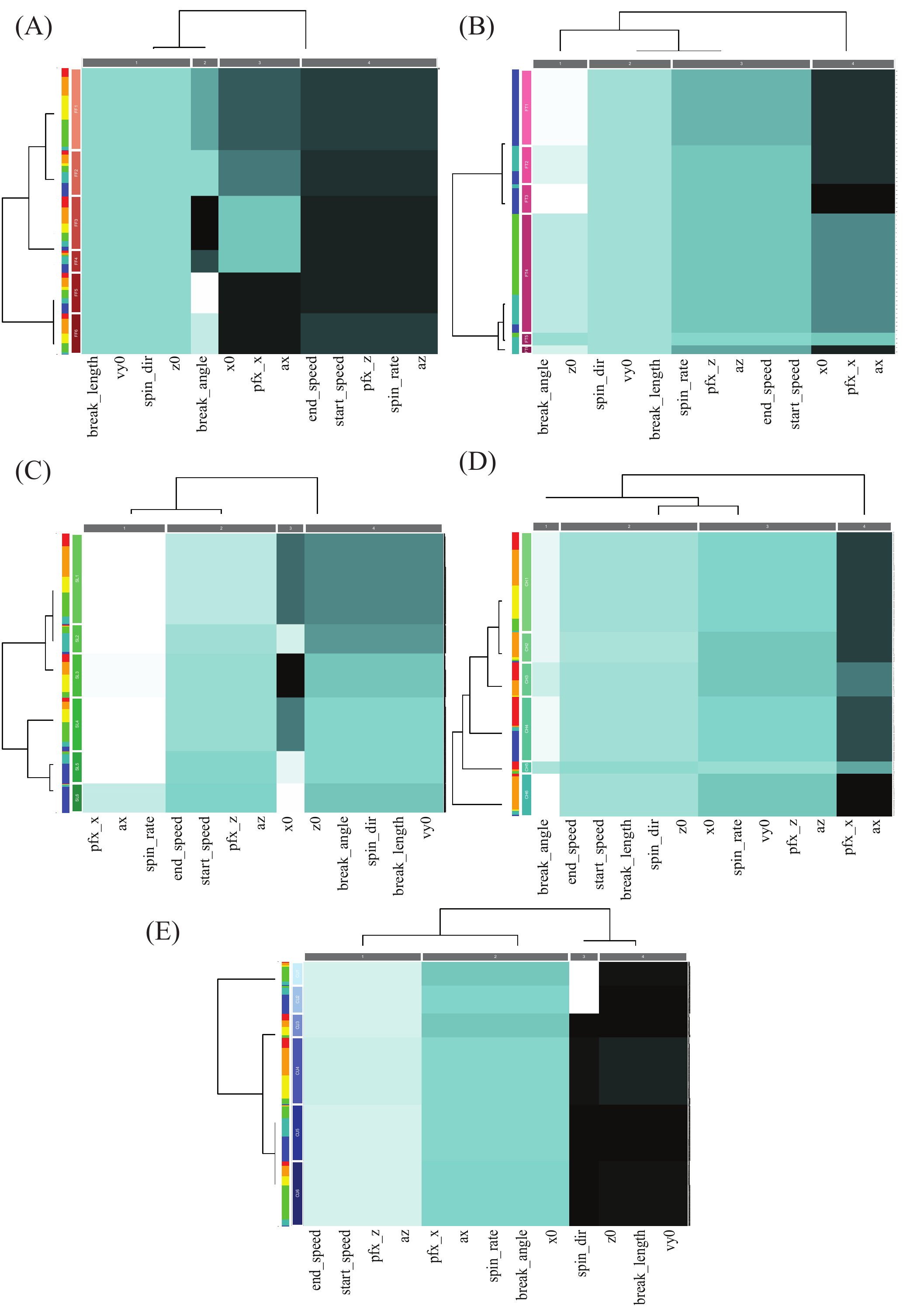}
	\caption{Heatmaps of pitch-type specific comparisons between Clayton Kershaw 2012-2014 seasons(baseline) with 2015-2017 seasons. (A) FF; (B) FT (C) SL; (D) CH; (E) CU.}
	\label{fig:figure6}
\end{figure}

The five pitch-type specific heatmaps are reported in Figure \ref{fig:figure6}. On the panel (A), all 4-seam fastballs (FF) are classified into 6 subtypes. The idiosyncratic characteristics of these 6 subtypes are displayed on the heatmap in individual and associative manners. Across these 6 subtypes, features: \{break\textunderscore length, spin-dir, vy0, z0 \} are kept constant at the lower end, features: \{Start\textunderscore speed, spin\textunderscore rate, pfx\textunderscore z, az, end\textunderscore speed\}, are also kept constant at the higher end. The idiosyncratic characteristics of the 6 pitch-subtypes are captured by the opposite-contrasting interactions between features: \{pfx\textunderscore x, ax, x0\} and feature: \{break\textunderscore angle\}.

When the release point of a fastball pitch has a large horizontal coordinate value \{x0\} companied with a large value of acceleration (ax), the fastball trajectory will have a large value of horizontal movement \{pfx\textunderscore x\} and a distinctively small value of \{break\textunderscore angle\}, as seen in subtypes FF5 and FF6. In a reverse fashion, small values of \{x0\} and \{ax\} will result into small \{pfx\textunderscore x\}, but large \{break\textunderscore angle\}, as seen in subtypes FF3 and FF4. And median values of \{x0\} and \{ax\} lead to median values of \{pfx\textunderscore x\} and \{break\textunderscore angle\}, as seen in subtypes FF1 and FF2.

Further the uneven proportions of color-coding across the 6 subtypes also bring out the information of how Kershaw distributes 4-seam fastball subtypes across the 6 seasons. One important observation is that the aforementioned 3 interacting characteristics: FF$1\&2$, FF$3\&4$ and FF$5\&6$ have been constantly and alternatively realized by Kershaw's fastballs. Such an observation might explain why his 4-seam fastball is still one of his most effective pitch-type in his pitching repertoire.

The 2-seam fastball (FT) was only added to Kershaw's repertoire in season 2015. The number of pitches in FT is much smaller than that in FF. However, its evolving patterns are evident and interesting to see, as shown in panel (B). The interacting relational patterns in panel (B) somehow evolves and extends between features: \{pfx\textunderscore x, ax, x0\} and \{break\textunderscore angle, z0\}. The larger values of \{pfx\textunderscore x, ax, x0\} couple with smaller values of \{break\textunderscore angle, z0\}, as seen in three subtypes FT1-3 and FT6, and the reverse pattern is seen in subtypes FT$4 \& 5$.

In summary, Kershaw created his 2-seam fastball by first reducing the vertical coordinate \{z0\} on the releasing point of his 4-seam fastball gesture. Such pitch subtypes: FT$4\&5$, are dominate in 2015-16 seasons. This modification on \{z0\} has gone further down in 2017 season, as seen in three subtypes FT$1\&2\&3$.

A slider in general is threw with its pitching gesture just like that of a fastball, but with a slightly reduced speed.  Slider is one of three major pitch-types in Kershaw's repertoire. Upon the panel (C), Kershaw's pitching mechanics for slider over the 6 seasons has retained the interacting patterns seen in pitching mechanics of fastball in panels (A) and (B), but in a slight scale: SL1 vs SL4 on seasons 2012-2015; SL2 vs SL5 on season 2016; and SL5 vs SL6 on season 2017.

Beyond the interacting pattern, like the shadow of fastball, we also see a gradual, but persistent pattern: values of \{Start\textunderscore speed, pfx\textunderscore z, az, end\textunderscore speed\} become slightly larger in values, while values of \{break\textunderscore angle, break\textunderscore length, spin-dir, vy0, z0\} become slightly smaller in values over the 6 seasons. The features: \{pfx\textunderscore x, ax, spin\textunderscore rate\} are kept constant in the lower end, except the in the subtype SL6, which dominated by 2017 season. The evident idiosyncratic characteristics of these 6 subtypes are characterized by values of feature: \{x0\}. This character also contains a persistent decreasing pattern over the 6 seasons.

\begin{figure}
	\centering
	\includegraphics[width=0.9\linewidth]{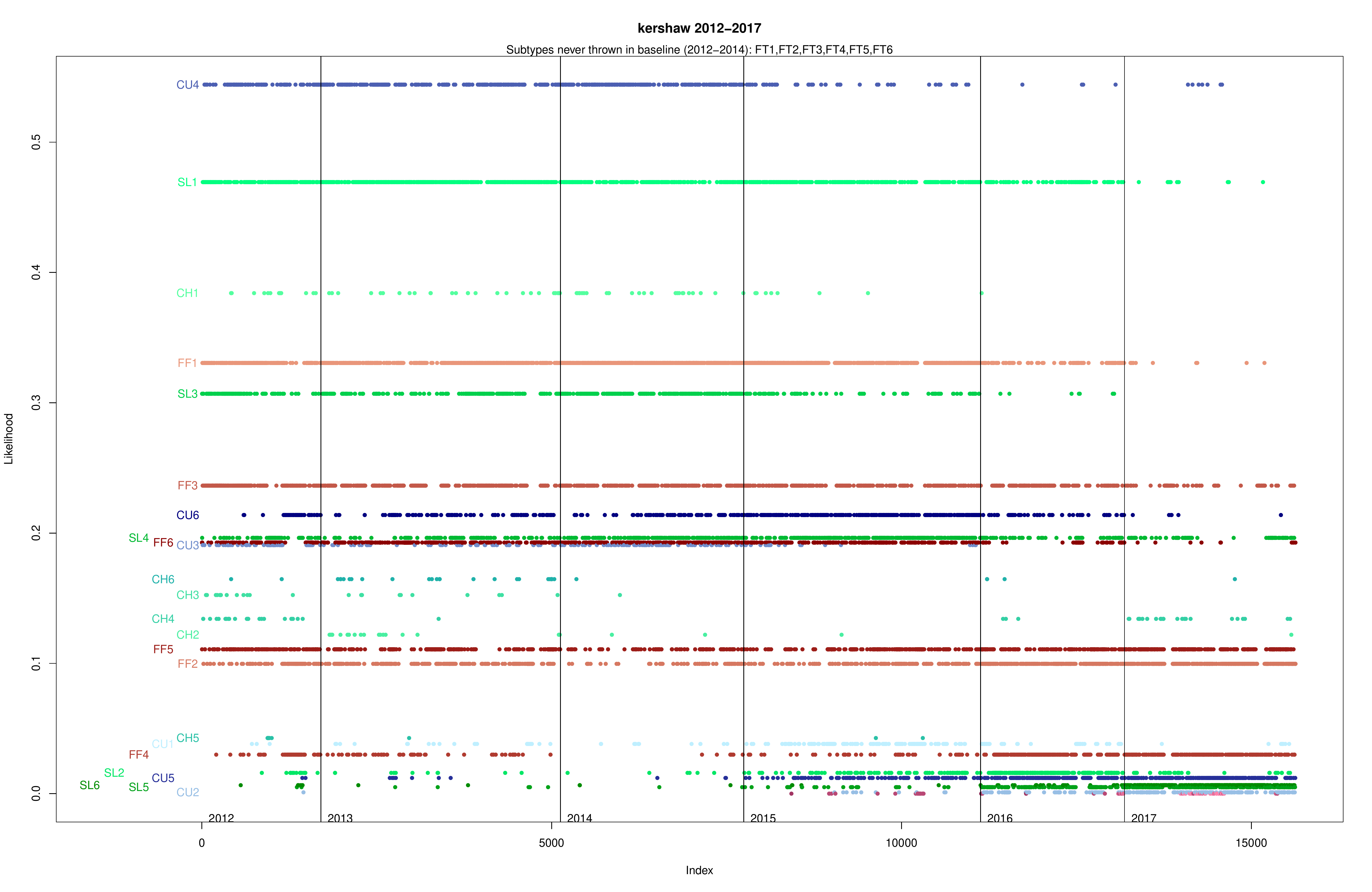}
	\caption{Pitch-subtypes based evolution for Clayton Kershaw 2012-2014 seasons (baseline) through 2015-2017 seasons.}
	\label{fig:figure7}
\end{figure}
Like slider, ideally a changeup is also thrown in a pitching gesture of fastball, but has a much reduced speed coupled with very diverse spin-direction. This diversity is carried out via many ways of holding a baseball when pitches a changeup. Different ways of holding induce different spin directions, and render different vertical and horizontal movements. So a changeup is disguised as a fastball in gesture, but has a drastically distinct and diverse movements. This is why it is an effective pitch-type for many MLB pitchers, but it is not a major pitch-type in Kershaw's repertoire.
\begin{figure}[hbtp]
	\centering
	\includegraphics[width=0.9\linewidth]{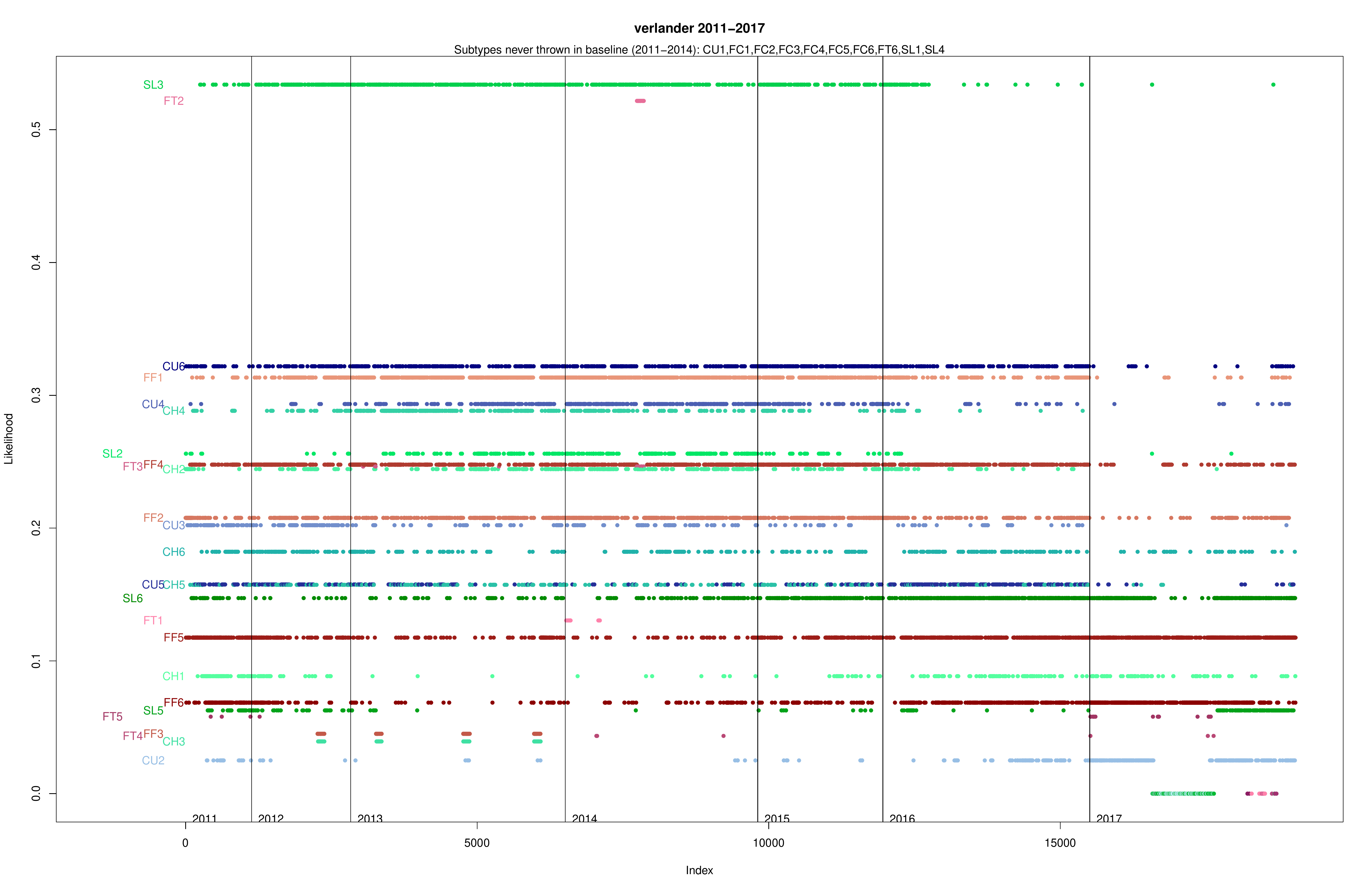}
	\caption{Pitch-subtypes based evolution for Justin Verlander 2011-2014 seasons (baseline) through 2015-2017 seasons.}
	\label{fig:figure8}
\end{figure}

In panel (D) of Figure \ref{fig:figure6}, the signature interacting pattern of Kershaw's fastball is seen only slightly in the heatmap for changeup (CH). It seems that Kershaw is on his way of phasing out this pitch-type altogether in the future season.

In terms of pitching gesture, Kershaw's curveball(CU) is thrown in a rather artificial way to create top-spinning.  It is not natural, so the speed is very low. Since its Magnus effect of going downward is further aided by gravity, its trajectory usually has a drastic, if not sudden, drop when a baseball is entering the home plate. That is why it is an effective pitch-type if a pitcher can introduce uncertainty into such a pitch, like Kershaw.

Curveball(CU) is one of Kershaw's most effective pitch-types in dealing with batters. The panel (E) for curveball (CU) clearly indicates that he throws two distinct subtypes of curveball with two opposite spin-direction: CU$1\&2$ with extremely low values of \{spin-dir\}, and CU$3\&4\&5\&6$ with extremely high values of \{spin-dir\}. That is, in general a batter has a hard time to guess which sides: right or left, his curveball will swerve to. This uncertainty makes his curveball well-known, even the height of release point of a pitch, i.e. feature \{z0\}, is kept detectably higher than that of all other pitch-types of his.  Further his curveball seemingly is narrowing into two subtypes: CU2 and CU5, in recent reasons. This might be an alarm sign of losing his effectiveness on this pitch-type.

Information contained in these five panels of Figure \ref{fig:figure6} is synthesized into a so-called likelihood plot, as shown in figure Figure \ref{fig:figure7}, with 2012-2014 being used as baseline. It is recalled that each pitch has a pitch-subtype ID and a likelihood value, which are derived from a pitch-type specific heatmap and its categorical pattern distribution in one of the five panels in Figure \ref{fig:figure6}. The likelihood plot displays each pitch's subtype-ID and likelihood vale with respect to its temporal coordinate upon the entire axis of 6 seasons. Such a graphic display collectively reveals nearly all pattern information of Kershaw's pitching mechanics, which evolves from 2012 through 2017. Many pitching subtypes go extinct, while many subtypes are created in later seasons. Also some subtypes even go into extinct, and then come back on again.

\begin{figure}[hbtp]
	\centering
	\includegraphics[width=0.9\linewidth]{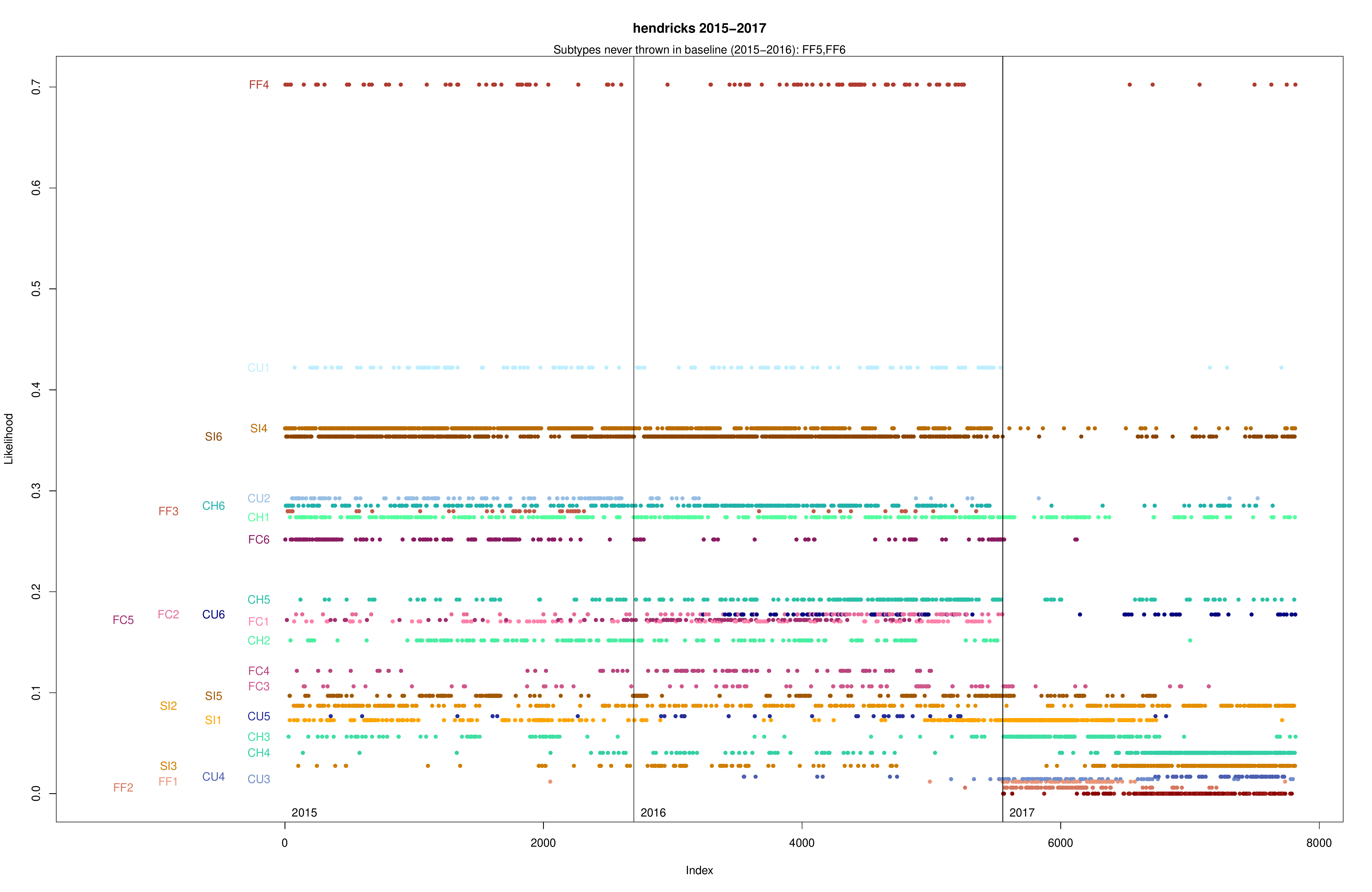}
	\caption{Pitch-subtypes based evolution for Kyle Hendricks 2015-2016 seasons (baseline) through 2017 season.}
	\label{fig:figure9}
\end{figure}

Two likelihood plots of Verlander and Hendricks are given in Figure \ref{fig:figure8} and Figure \ref{fig:figure9}, respectively. The likelihood plot with 2011-2014 being taken as the baseline seasons, shows that Verlander has 5 pitch-types: \{FF, FT, SL, CH, CU\}. In fact he has 6 pitch-types in the PITCHf/x database, including the pitch-type: Cuter (FC), which was created only recently way after 2014 season. He also creates one new curveball subtype, one 2-seam fastball subtype and two slider subtypes. In the Figure \ref{fig:figure3}, Verlander-2013 is located in the ``a-row" branch, while the rest of 7 pitcher-seasons are in ``b-row" branch. This separation can be somehow seen in this likelihood plot as well.

The likelihood plot of Hendricks' pitching mechanics on three seasons: 2015-2017 is shown in Figure \ref{fig:figure9}. The 2015-2016 seasons constitute the baseline and there are five pitch-types: FF, FC, SI(sinker), CH, CU. This likelihood plot shows a visible evident change from 2016 to 2017 season.  Many subtypes are nearly extinct, while many sparsely used subtypes in the baseline seasons are heavily used in the 2017 season. This evident change indeed coincided with his injury at early part of 2017 season.

\section{Conclusion}
Though MLB pitcher's pitching mechanics are complex and somehow mysterious, our data-driven computing and graphic displays are able to lift the veil to certain extents. Information of several aspects of pitching mechanics is genuinely extracted and then explicitly represented for systemic comparison among all 24 pitchers, and for individual pitcher's idiosyncratic evolution across multiple seasons. Understanding of computed information via a graphic display as a platform seems to be able to flow smoothly. Nonetheless that is just the beginning of a new era into this 150 year old sport.

The systemic comparison based on a heatmap, derived from mutual conditional entropy matrices of all pitcher-seasons, not only enables us to see who are close to whom, and far away from whom among MLB pitchers, but also pinpoints which features can be accounted for their differences in pitching mechanics. Equally importantly, the idiosyncratic evolution of pitching subtypes in a pitcher's repertoire shows the history of how this pitcher maintains, creates or gives up pitching subtypes across seasons, along which his career peaks, slides or declines.

In this big-data era, it is critical to develop computing platforms in order to properly address the fundamental question: what and where is information contained in data? In this paper, the discovery of the universal block patterns, which embed with all involving physical laws on all mutual conditional entropy matrices, is an answer to this question. It also is a validity check of our data-driven computations.

However we have to admit that results and understanding presented here fulfill only a small part of our primary objective that our computational endeavors intend to achieve upon PITCHf/x database. Our primary objective is to use our data-driven computing and graphic displays as platforms for all people to explore and discover information by themselves, and building and constructing understanding and knowledge from the database for themselves. This whole scale objective can't be achieved in a classic format of a published paper because of its rather limited space. Therefore it becomes necessary to create a website to facilitate all potential explorations and discoveries possibly accomplished by all curious minds. This is one way to accommodate all possible relevant information about pitching mechanics contained in this Big-Data era.

People, who has a career centering around or in baseball pitching, are able to find the landscape of pitchers and their differences in pitching mechanical factors for management purposes as well as for self-evaluations linked to pitching wellbeing.  People, who are interested in baseball as sport fans, are able to find deeper insights into favorite or adversarial pitchers' characteristics in pitching mechanics.  People, who are interested in data-driven computing or data science in general, are to find inspirations from resolutions of various issues on high dimensional point-cloud geometries, and, at the same time, to discover a new direction of research into biomechanics of pitching.

Particularly young people, who are guided by their own curiosity in baseball, are able to wonder around the new data-driven physical and aerodynamic laws embedded within pitching mechanics, and to appreciate their tangled complexity. The immediate sense of achievement to them is being able to go far beyond the more than 150-year-old Boxscores. Such an educational merit could be way bigger than the total sum of aforementioned ones. Young people likely rediscover baseball games with brand new insights, and then perceive this sport from totally different perspectives that their parents', grand-parents' generations never have imagined.

\vspace{0.5in}

\section*{\bf Ethics:} No ethical approvals are needed in this study.  All measurements pertaining to study subjects are available from two public websites.

\section*{\bf Data Accessibility:} The pitching data is available in PITCHf/x database belonging to Major League Baseball via \url{http://gd2.mlb.com/components/game/mlb/}.

\section*{\bf Competing Interests:} We have no competing interests.

\section*{\bf Author's contributions:} H.F. designed the study. K. F. collected all data for analysis. H.F., K. F.and T.R analyzed the data. H.F, K. F., T.R. C-J. H., and B. C. interpreted the results. H.F wrote the manuscript. H.F, K. F., T.R. C-J. H., and B. C. edited the manuscript.  All authors gave final approval for publication.

\section*{\bf Funding:} No financial funding received for this study.

\section*{\bf Acknowledgement:} We thank MLBAM for making the PITCHf/x data available.

\end{document}